# Energy-efficient tunable silicon photonic micro-resonator with graphene transparent nano-heaters

Longhai Yu[1], Yaocheng Shi[1], Daoxin Dai[1,] * and Sailing He[1, 2,] *

[1]Centre for Optical and Electromagnetic Research, JORCEP, State Key Laboratory for Modern Optical Instrumentation, Zhejiang Provincial Key Laboratory for Sensing Technologies, Zhejiang University, Zijingang Campus, Hangzhou 310058, China.

[2]Department of Electromagnetic Engineering, School of Electrical Engineering, Royal Institute of Technology, S-100 44 Stockholm, Sweden.

*E-mail: dxdai@zju.edu.cn, sailing@kth.se

Abstract: Thermally-tuning silicon micro-cavities are versatile and beneficial elements in low-cost large-scale photonic integrated circuits (PICs). Traditional metal heaters used for thermal tuning in silicon micro-cavities usually need a thick $SiO_2$ upper-cladding layer, which will introduce some disadvantages including low response speed, low heating efficiency, low achievable temperature and complicated fabrication processes. In this paper, we propose and experimentally demonstrate thermally-tuning silicon micro-disk resonators by introducing graphene transparent nano-heaters, which contacts the silicon core directly without any isolator layer. This makes the graphene transparent nano-heater potentially to have excellent performances in terms of the heating efficiency, the temporal response and the achievable temperature. It is also shown that the graphene nano-heater is convenient to be used in ultrasmall photonic integrated devices due to the single-atom thickness and excellent flexibility of graphene. Both experiments and simulations imply that the present graphene transparent nano-heater is promising for thermally-tuning nanophotonic integrated devices for e.g. optical modulating, optical filtering/switching, etc.

Keywords: graphene; transparent nano-heater; silicon micro-disk resonator; thermal tuning

1. Introduction

Silicon photonics is one of the most promising technologies to realize low-cost large-scale photonic integrated circuits (PICs) because of the ability for ultra-sharp bends and the compatibility with CMOS fabrication processes.[1, 2] Among various silicon photonic integrated devices, an optical micro-cavity is well known as a versatile element to realize many functionality components, including optical filters/switches,[3, 4] optical modulators[5] and optical sensors.[6] Since silicon has a large heat conductivity (~149W/m·K) and a large thermo-optical (TO) coefficient (~1.8×10$^{-4}$/K at the wavelength of 1.55μm),[7] it is possible and beneficial to achieve efficient thermal tuning in silicon micro-cavities. Traditionally, metal heaters are usually used for thermally tuning silicon micro-cavities, and a thick $SiO_2$ upper-cladding layer is required between the metal heater and the silicon core to isolate metal absorption. However, this thick $SiO_2$ upper-cladding layer will introduce some disadvantages due to its poor heat conductivity, e.g., low response speed, low heating efficiency and low temperature increase. Moreover, silicon micro-cavities are desired to have small size for large-scale integration, in which case the fabrication becomes difficult for traditional metal heaters. Therefore, a novel and efficient heating approach for thermally-tuning silicon micro-cavities are desired very much.

Graphene has been extensively investigated all over the world because this two-dimensional (2D) sheet has many extraordinary properties,[8, 9, 10] such as 0.34nm-thickness for monolayer graphene,[11] broadband absorption of ~2.3% per layer for vertically incident light,[8, 9] carrier mobility as high as 200,000cm$^2$/V·s at room temperature (RT),[12, 13] a "minimum" conductivity of ~4$e^2$/h even when the carrier concentration tends to zero,[8, 10] a high value of optical damage threshold and excellent mechanical stability.[9] Recent studies also suggest that graphene is expected to have high intrinsic thermal conductivity owing to the long-wavelength phonon transport in its 2D crystal lattices,[14, 15] and an experiment value of up to 5300W/m·K at RT has confirmed this conclusion.[16] Excellent thermal property of graphene coupled with its remarkable optoelectronic properties enables many potential applications of graphene in thermal management,[17, 18] such as efficient heat spreaders in electronic and photonic devices,[19, 20] transparent and flexible heaters,[21, 22, 23] etc.[24, 25]

In this paper, we experimentally demonstrate a thermally-tuning silicon micro-disk resonator with graphene transparent nano-heaters for the first time. The graphene nano-heater is designed to contact the silicon core directly without the thick $SiO_2$ upper-cladding layer required for traditional metal heaters, which makes it possible to achieve efficient thermal tuning, high temporal response as well as simple fabrication processes. Due to the single-atom thickness and excellent flexibility of graphene, it is convenient for the transparent nano-heater to be patterned in a nano-scale shape with CMOS-compatible fabrication processes. This is very useful to small-size nanophotonic integrated devices that need thermal management, i.e., the thermally-tuning micro-disk resonator in our case. Here we characterize the thermally-tuning micro-disk resonator through a series of experiments and simulations, and show that the graphene transparent nano-heater is a novel, beneficial and promising heating method for nanophotonic integrated circuits.

2. Materials and Methods

The three-dimensional (3D) schematic illustration of the thermally-tuning silicon micro-disk with a graphene transparent nano-heater is shown in Figure 1a. The fabrication starts with a commercial SOI wafer with a 250nm-thick top silicon layer and a $SiO_2$ buried oxide (BOX) layer of 3μm. We use electron beam lithography (EBL) and inductive coupling plasma (ICP) etching processes to form the submicron-patterns. Then a second EBL process and a lift-off process are carried out to make a couple of 100nm-thick titanium metal contacts on top of the $SiO_2$ BOX layer. A monolayer graphene sheet, grown by chemical vapor deposition (CVD) method, is wet-transferred onto the whole SOI chip without any alignment.[26, 27] A third EBL process followed by an oxygen plasma etching process is then utilized to pattern the graphene sheet forming the nano-heater as well as the connecting bridges that connect the nano-heater and the metal contacts.[26] As shown in Figure 1b, the graphene nano-heater has an annulus part on the micro-disk and two straight parts extending out of the micro-disk. The radii of the micro-disk and the annulus nano-heater are about $r_d$=5μm and $r_h$=3.4μm respectively. And the widths of the graphene nano-heater are about $w_h$=1.2μm for the annulus part and $w_{ch}$=2μm for the straight part. These designing parameters are used to keep the annulus graphene nano-heater being laid outside the region of the whispering-gallery mode (WGM) of the micro-disk resonator (see the bottom of Figure

1b),[28] which avoids introducing extra loss by the graphene sheet on the silicon core. Thus a transparent nano-heater is achieved.

Figure 1c shows the microscope images of the fabricated thermally-tuning silicon micro-disk resonators. To protect the patterned graphene from damage or contamination, the photoresist for patterning is not removed during the experiment. Since the guided mode is mainly confined to the silicon core (see the bottom of Figure 1b), the remaining photoresist on the graphene sheet will not influence thermal behaviors of the devices notably.[23] In Figure 1d, we show the scanning electron microscope (SEM) images of the micro-disk resonator covered by the patterned graphene nano-heater. It can be seen that the graphene sheet flexibly crosses the step from the top surface of the $SiO_2$ BOX layer to that of the silicon core. Therefore, the present graphene nano-heater is available for complex surfaces with non-planar nanostructures even though there is no $SiO_2$ upper-cladding layer.

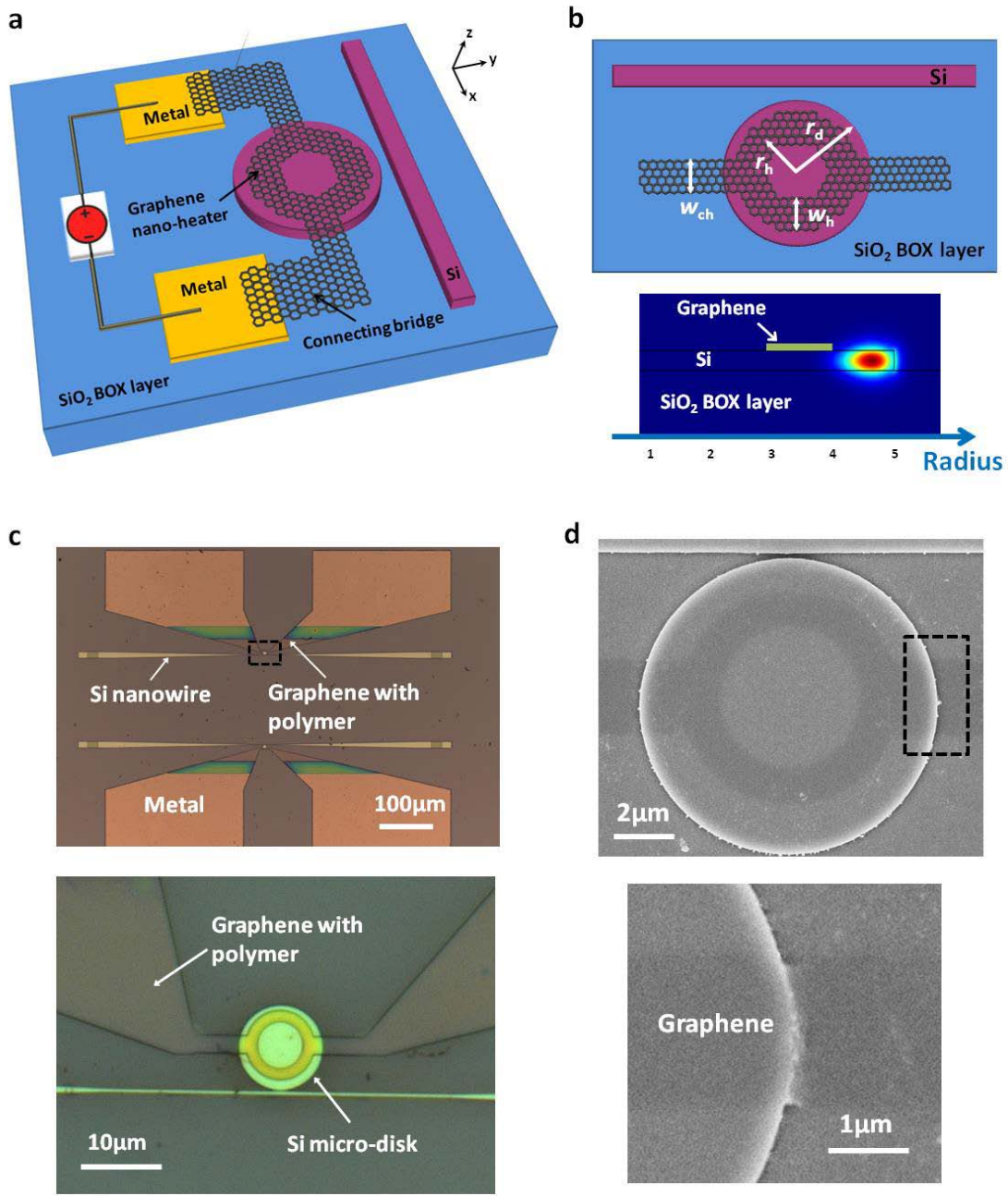

**Figure 1** Thermally-tuning silicon micro-disk resonator with a graphene transparent nano-heater. **(a)** Three-dimensional schematic illustration of a thermally-tuning silicon micro-disk with a graphene transparent nano-heater. The graphene nano-heater has an annulus part on the micro-disk and two straight parts extending out of the micro-disk. The polymer on the graphene is not shown. **(b)** Top, top view for the micro-disk resonator with a graphene nano-heater. The radii of the micro-disk and the annulus nano-heater are about $r_d$=5μm and $r_h$=3.4μm respectively. The widths of the graphene nano-heater are about $w_h$=1.2μm for the annulus part and $w_{ch}$=2μm for the straight part. Bottom, cross-section view for the whispering-gallery mode (WGM) of the micro-disk resonator. The annulus

graphene nano-heater on the micro-disk is shown in a green line. **(c)** Top, top-view microscope image of the fabricated thermally-tuning micro-disk resonator (Scale bar: 100µm). Bottom, zoom-in view for the black dash square in the top image (Scale bar: 10µm). The graphene sheet is covered by a polymer film. **(d)** Top, top-view scanning electron microscope (SEM) image of the micro-disk resonator covered by the patterned graphene nano-heater (Scale bar: 2µm). Bottom, zoom-in view for the black dash square in the top image (Scale bar: 1µm). The graphene sheet flexibly crosses the step from the top surface of the SiO$_2$ BOX layer to that of the silicon core. The polymer on the graphene sheet is removed.

3. Results and Discussion

As is well known, the spectral response of a micro-disk resonator can be described as[29]

$$P_{out}(\lambda) = P_{in} \left[ \frac{t - e^{-i2\pi r_d (\frac{2\pi n_e}{\lambda} - \alpha)}}{1 - t e^{-i2\pi r_d (\frac{2\pi n_e}{\lambda} - \alpha)}} \right]^2, \quad (1)$$

where $P_{in}$ and $P_{out}$ are the input and output powers, $\lambda$ is the wavelength of the probe light, $t$ is the transmission ratio in the coupling region of the micro-disk, $\alpha$ is the attenuation coefficient, and $n_e$ is the effective refractive index of the WGM. For the thermally-tuning micro-disk resonator in our case, when a heating power is loaded in the graphene sheet through the two metal contacts, heat will be generated in the graphene nano-heater because of the large resistance in this region. The heat is then transferred directly from the graphene nano-heater to the silicon micro-disk below, and this causes the temperature increase of the silicon. Due to the positive TO coefficient of silicon (i.e., ~1.8×10$^{-4}$/K), the effective refractive index of the micro-disk resonator $n_e$ increases. As indicated in Equation (1), a phase variation is then introduced in the resonator, and consequently the spectral response of the micro-disk has a redshift thermally.

Figure 2a shows the experiment result for the dynamic spectral responses of our fabricated thermally-tuning micro-disk resonator. The resonant wavelength of the micro-disk resonator has a redshift of ~5nm when the heating power $P_{heating}$ varies from 0mW to 10.5mW. Then the heating efficiency, which is defined as the shift of the resonant wavelength with unit heating-power consumption, can be obtained as about $\eta_d$=0.474nm/mW. This value indicates the conversion efficiency from the electrical heating power to the temperature change of the silicon micro-disk, which includes a series of processes, e.g., heat generation in the graphene nano-heater, heat transfer from the graphene sheet to the silicon core, and heat delivery in the 5μm-radius micro-disk. The temporal response of the thermally-tuning micro-disk resonator is also measured. We can get the 90% rising and decaying times of thermal tuning are ~12.8μs and ~8.8μs respectively. In our device, the graphene nano-heater contacts directly with the silicon core, so the time for heat transfer from graphene to silicon can be neglected, and the temporal response here mainly depends on the heat delivery in the silicon micro-disk.

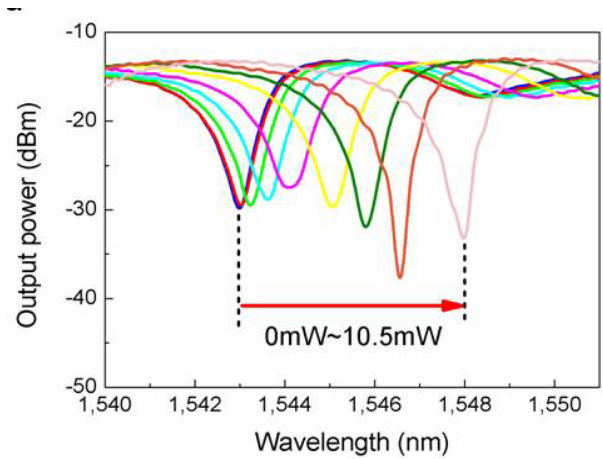

**Figure 2** Spectral responses of the thermally-tuning micro-disk resonator as the heating power $P_{heating}$ varies from 0mW to 10.5mW. The heating efficiency is about $\eta_d$=0.474nm/mW.

In order to simulate the thermal behaviors of the present thermally-tuning micro-disk resonator, we use a 3D finite element method (FEM) tool to solve the Poisson Equation. In the simulation, we assume that the electrical energy for heating is completely converted to thermal energy, and the thermal energy is distributed averagely in the graphene nano-heater. When the heating power in unit area of the graphene nano-heater is about ~0.2mW/μm$^2$, we can obtain the temperature distribution of the silicon micro-disk resonator in Figure 3a. It shows that the regions around the graphene nano-heater have higher temperature than the regions far away from the graphene nano-heater, and the highest temperature (i.e., ~431.8K) appears in the connecting regions of the annulus and straight nano-heaters. To see clearly the heat delivery in the heated micro-disk, we show the x, y and z components of the total heat flux in Figure 3b-3d. The figures indicate that the shape of the graphene nano-heater (i.e., $w_{ch}$, $w_h$ and $r_h$) influences the result significantly, and it is possible to increase the heating efficiency by further optimizing the shape (see below). In our device, the annulus graphene nano-heater is designed to be just near the guided mode (WGM) around the micro-disk resonator (see the bottom of Figure 1b). This is helpful to realize an efficient increase for the effective refractive index of the WGM even though the temperature distribution is not average in the heated micro-disk.

To derive a theoretical heating efficiency, we can calculate the effective temperature and the corresponding effective refractive index $n_e$ for the WGM according to its field distribution in the micro-disk (see the bottom of Figure 1b). Then by using Equation (1), the spectral response of the micro-disk is obtained, and the theoretical heating efficiency can be derived as about $\eta_t$=1.25nm/mW. This value is higher than the measurement result in our experiment (i.e., ~0.474nm/mW). The main reason is because the conversion efficiency from the electrical energy for heating to the thermal energy in the graphene nano-heater can never be ideally 100%. The heat generated in the graphene nano-heater is not only transferred to the silicon core but also delivered to the connecting bridges due to the high intrinsic heat conductivity of graphene.[16] Then the thermal energy in the connecting bridges will be wasted to heat the SiO$_2$ BOX layer below them. Therefore, the assumption in the simulation cannot be realized in our experiment. Another reason is the quality degradation of the graphene sheet during the fabrication process. It is observed that the wet-transferred graphene sheet has some contamination and cracks, which affects the heat transfer from graphene to silicon and reduces the heating efficiency. With the simulation, we can also get the temporal response of the thermally-tuning micro-disk resonator. The theoretical 90% rising and decaying times are both ~8.3μs, which are slightly faster than the measurement results (i.e. ~12.8μs for the rising process and ~8.8μs for the decaying process). This results from the quality degradation of the graphene nano-heater as well, and the influence to the heating process (rising process) is larger than that to the cooling process (decaying process).

To make a comparison, we simulate a thermally-tuning silicon micro-disk resonator with a traditional metal heater by using the FEM tool. The metal heater has the same shape as the present graphene nano-heater, and the SiO$_2$ upper-cladding layer between the metal heater and the silicon core is about ~1μm. From the simulation, we can get the heating efficiency of about $\eta_m$=0.94nm/mW, which is about 25% smaller than that with the graphene nano-heater (i.e. $\eta_t$=1.25nm/mW). The reduced heating efficiency for the metal heater is owing to the addition of the thick SiO$_2$ upper-cladding layer, which increases the heating volume significantly. The 90% rising and decaying times with the metal heater are also obtained as ~12.8μs. This value is about 54% longer than the result with the graphene nano-heater (i.e., ~8.3μs) due to the poor heat conductivity of the SiO$_2$ upper-cladding region. By using metal heaters, it takes some time to deliver heat from the metal to the silicon core, while this problem is removed with graphene nano-heaters because graphene nano-heaters contact the silicon directly. Furthermore, we find that the temperature of the silicon core is much lower than that of the heater (e.g., $\Delta T=T_{metal}-T_{silicon}$=~89K when the heating power in unit area of metal is ~0.2mW/μm$^2$) for the case of metal heaters. In contrast, when using graphene nano-heaters, the temperature of the silicon core is almost as same as that of the heater. Therefore, graphene nano-heaters are beneficial to achieve a higher operation temperature and a larger tunable range. The comparison above implies that graphene nano-heaters are possible to achieve better performance than traditional metal heaters in terms of heating efficiency, temporal response and maximal operation temperature for thermally-tuning nanophotonic integrated devices.

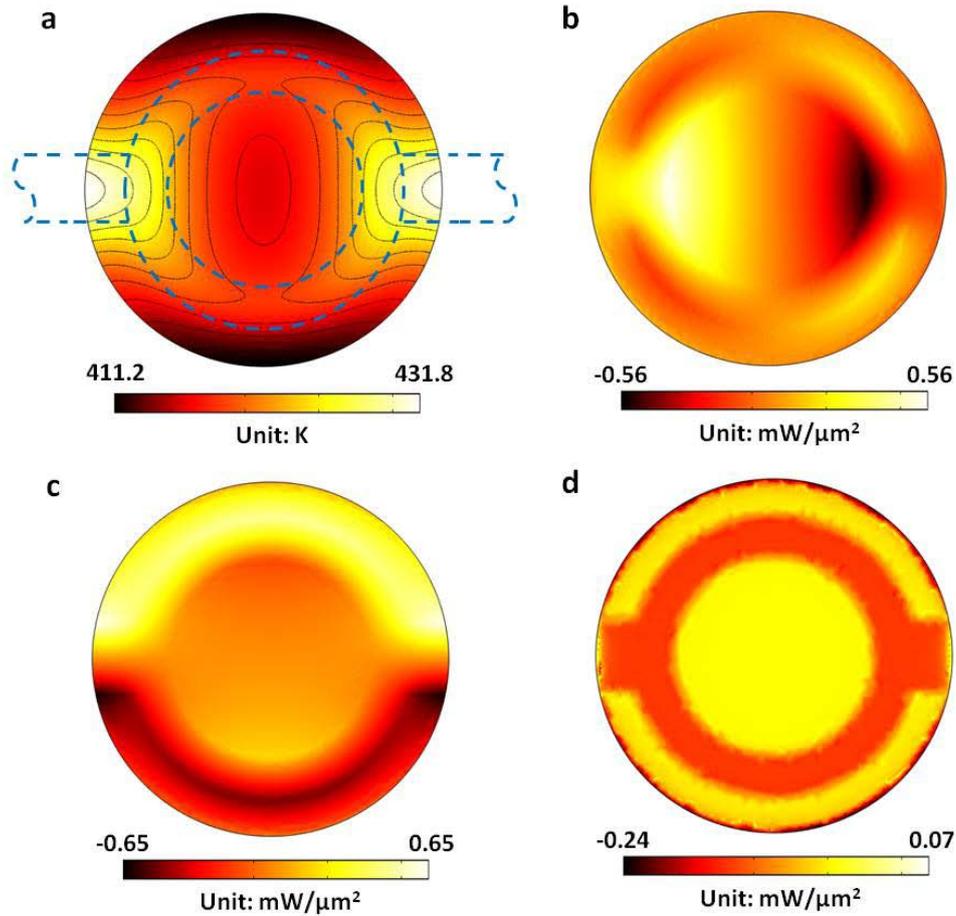

**Figure 3** Simulation for the thermally-tuning micro-disk resonator with a graphene transparent nano-heater by solving the Poisson Equation with a 3D finite element method (FEM) tool. The heating power loaded in unit area of the graphene nano-heater is ~0.2mW/µm². **(a)** Temperature distribution of the heated micro-disk resonator. The outlines of the graphene nano-heater are shown in blue dash lines. Room temperature is 300K. **(b)** x component, **(c)** y component and **(d)** z component of the total heat flux in the heated micro-disk resonator. The coordinate axis is shown in Figure 1a.

The experiment and simulation results above have shown that the present graphene transparent nano-heater provides efficient thermal tuning for silicon micro-disk resonators. It is also noted that the heating efficiency and temporal response can be enhanced with great potential. First, we can adjust the shape of the graphene nano-heater to optimize the temperature distribution in the micro-disk, which is beneficial to realize higher heating efficiency and faster temporal response. For example, Figure 4 shows the heating efficiency of the graphene nano-heater with varying widths for its annulus part ($w_h$) and straight part ($w_{ch}$). Considering the propagation loss, we keep the outer edge of the annulus graphene nano-heater (i.e., $r_h+w_h/2$ in Figure 1b) constant as ~4µm for 5µm micro-disk resonator. The results of our experiment devices are shown as stars in the figure. We find that by increasing $w_h$ and decreasing $w_{ch}$, the heating efficiencies can be as high as ~1.55nm/mW for 5µm-radius thermally-tuning micro-disk resonators respectively. However, a larger $w_h$ can reduce the resistance of the annulus graphene nano-heater and limit heat generation in this region, while a smaller $w_{ch}$ may decrease the current damage threshold and the maximal operation current of the graphene nano-heater.[33] Therefore, there is a trade-off among the heating efficiency, resistance and maximal operation current in the process of designing the graphene nano-heater. In addition, the resistance of the graphene nano-heater can also be optimized. Reducing the contact resistance as well as increasing the resistance of the annulus part helps to cut the energy waste and achieve improved thermal tuning. What is more, the metal contacts can be placed closer to the graphene nano-heater so that the area of the graphene connecting bridges is shrunk to reduce heat spreading. The simulation shows that the thick $SiO_2$ BOX layer is such a large heat sink that it limits the heating efficiency significantly, especially when the straight graphene nano-heater and the connecting bridges contact the $SiO_2$ BOX layer directly in our case. To overcome this problem, one can suspend the graphene sheet[6, 13] or insert a layer with low heat conductivity (e.g., polymer) between the graphene sheet and the substrate to prevent the heat from dissipating. Another important approach to improve the present heating efficiency and temporal response is to further enhance the graphene quality by optimizing the fabrication processes.

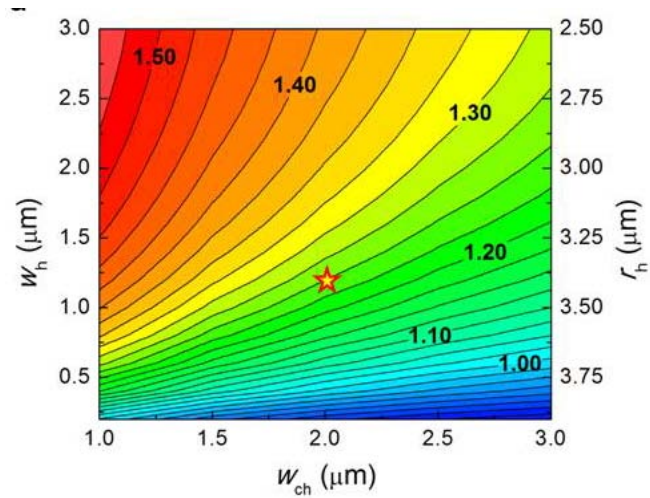

**Figure 4** Theoretical heating efficiency of the thermally-tuning micro-disk resonator with a graphene transparent nano-heater. The widths of the graphene nano-heater in the straight part ($w_{ch}$) and annulus part ($w_h$) are varied. The outer edge of the annulus graphene nano-heater (i.e., $r_h+w_h/2$) keeps constant as ~4μm for the 5μm-radius micro-disk resonator. The corresponding radius of the annulus nano-heater $r_h$ is also shown. The parameters used in our experiments are indicated as stars.

4. Conclusion

In summary, a thermally-tuning silicon micro-disk resonator with a graphene transparent nano-heater is demonstrated for the first time. By optimizing the shape of the graphene nano-heater, it will introduce little extra loss even when the graphene sheet contacts directly with the silicon core. Both experiments and simulations have shown that the present graphene transparent nano-heater could achieve good performance in terms of heating efficiency, temporal response, maximal operation temperature as well as fabrication process in nano-scale devices. It is indicated that this novel heating method based on graphene has great promise for many applications in thermally-tuning nanophotonic integrated devices, such as optical modulating, optical filtering/switching and so on.

5. Acknowledgements

This work was partially supported by the National Natural Science Foundation of China (No. 91233208, 61422510), the National High Technology Research and Development Program (863) of China (No. 2012AA012201), and the Program of Zhejiang Leading Team of Science and Technology Innovation.